\documentclass
    [    aip, jcp,
	, superscriptaddress
	, showpacs
	, floatfix
        , citeautoscript 
	 , twocolumn , reprint
    ]
    {revtex4-1}

\usepackage{graphicx}
\usepackage{epsfig,amsmath,amsfonts,amssymb,wasysym,xcolor} 
\usepackage{soul}
\usepackage{textcomp}
\usepackage{hyperref}

\newcommand{\of}[1]{\ensuremath{\left(#1\right)}}
\newcommand{\abs}[1]{\ensuremath{\left\vert#1\right\vert}}
\newcommand{\fd}[1]{\ensuremath{\left[#1\right]}}
\newcommand{\set}[1]{\ensuremath{\left\lbrace#1\right\rbrace}}
\newcommand{\thermal}[1]{\ensuremath{\left\langle#1\right\rangle}}

\newcommand{\tcb}[1][]{\ensuremath{T_c^{(s)}}}
\newcommand{\Op}[1][]{\ensuremath{\phi_{#1}}}
\newcommand{\T}[1][]{\ensuremath{\mathcal{T}_{#1}}}
\newcommand{\ccf}[1][]{\ensuremath{f_{C#1}}}
\newcommand{\sfeos}[1][]{\ensuremath{F_{#1}}}
\newcommand{\sff}[1][]{\ensuremath{\vartheta_{#1}}}
\newcommand{\sfp}[1][]{\ensuremath{\theta^{#1}}}

\newcommand{\hb}[1][]{\ensuremath{h_{b#1}}}

\newcommand{\bsec}[1][]{\ensuremath{B_{2#1}}}
\newcommand{\w}[1][]{\ensuremath{W}}

\newcommand{\m}[1][]{\ensuremath{\omega_{#1}}}

\newcommand{\x}[1][]{\ensuremath{\mathbf{r}_{#1}}}
\newcommand{\q}[1][]{\ensuremath{\mathbf{q}_{#1}}}
\newcommand{\pot}[1][]{\ensuremath{U_{#1}}}

\newcommand{\sgn}[1][]{\ensuremath{\mathop{sgn}}}
\newcommand{\ft}[1]{\ensuremath{\hat{#1	}}}

\newcommand{\eref}[1]{{Eq.~\eqref{#1}}}
\newcommand{\fref}[1]{{Fig.~\ref{#1}}}

\begin{document}
\title{Structure and aggregation of colloids immersed in critical solvents}
\date{\today}
\author{T. F. Mohry}
\email{mohry@is.mpg.de}
\affiliation{Max-Planck-Institut f{\"u}r Intelligente Systeme,
  Heisenbergstra{\ss}e 3, 70569 Stuttgart, Germany}
\affiliation{Universit{\"a}t Stuttgart, Institut f{\"u}r Theoretische und Angewandte Physik,
  Pfaffenwaldring 57, 70569 Stuttgart, Germany}
\author{A. Macio{\l}ek}
\email{maciolek@is.mpg.de}
\affiliation{Max-Planck-Institut f{\"u}r Intelligente Systeme,
  Heisenbergstra{\ss}e 3, 70569 Stuttgart, Germany}
\affiliation{Universit{\"a}t Stuttgart, Institut f{\"u}r Theoretische und Angewandte Physik,
  Pfaffenwaldring 57, 70569 Stuttgart, Germany}
\affiliation{Institute of Physical Chemistry, Polish Academy of Sciences,
  Kasprzaka 44/52, PL-01-224 Warsaw, Poland}
\author{S. Dietrich}
\email{dietrich@is.mpg.de}
\affiliation{Max-Planck-Institut f{\"u}r Intelligente Systeme,
  Heisenbergstra{\ss}e 3, 70569 Stuttgart, Germany}
\affiliation{Universit{\"a}t Stuttgart, Institut f{\"u}r Theoretische und Angewandte Physik,
  Pfaffenwaldring 57, 70569 Stuttgart, Germany}
\begin{abstract}
We consider an ensemble of spherical colloidal particles immersed
in a near-critical solvent such as a binary liquid mixture close to
its critical demixing point. The emerging long-ranged fluctuations of
the corresponding order parameter of the solvent drive
the divergence of the correlation length.
Spatial confinements of these critical fluctuations by colloidal solute
particles, acting as cavities in the fluctuating
medium, restrict and modify the fluctuation spectrum in a way which 
depends on their relative configuration.
This results in effective, so-called
critical Casimir forces (CCFs) acting on the confining surfaces.
Using the available knowledge about CCFs we study
the structure and stability of such colloidal suspensions by
employing an approach in terms of effective, {\it one}-component
colloidal systems. Applying the approximation of pairwise additive 
CCFs we calculate the radial distribution function of
the colloids, which is experimentally accessible.
We analyze colloidal aggregation due to CCFs
and thus allude to previous experimental studies which are still
under debate.
\end{abstract}
\pacs{61.20.Gy, 64.60.fd, 64.70.pv,64.75.Xc, 82.70.Dd}
\maketitle
\section{Introduction}

In colloidal suspensions with near-critical solvents, colloidal
particles acting as cavities impose boundary conditions
at the colloid surfaces for the fluctuating order-parameter (OP)
of the solvent and perturb the OP field on the length scale of the
bulk correlation length 
$\xi\of{t=\of{T-\tcb}/\tcb\to 0}\sim \abs{t}^{-\nu}$, which
diverges upon approaching the bulk critical temperature \tcb{}
of the {\it s}olvent \cite{Fisher-et1978}; $\nu$ is a standard
critical bulk exponent.
These modifications of the OP and the restrictions of its fluctuation spectrum
result in an effective force acting between the colloids, known as the 
so-called critical Casimir force (CCF) \cite{Fisher-et1978}.
For equal boundary conditions, i.e., equal preference of all colloids
for one of the two species of the binary solvent, the CCF is attractive
and can lead to reversible aggregation phenomena. 
Aggregation phenomena are of considerable interest due to their relevance
for the stability of colloidal suspensions.

In our preceding paper \cite{MMD_part1} we have investigated the
phase behavior of  colloidal suspensions with near-critical
solvents and have explored the possibility of a phase separation 
into two thermodynamic phases, one being rich and the other
being poor in colloidal particles.
If the effective attraction potential between colloidal particles 
is sufficiently strong, a condensation of colloids into a stable 
{\it equilibrium} liquid phase may be preempted (on the time scale
of observation) by the formation
of {\it non-equilibrium} aggregates in
which the colloidal particle stick together. In general, these aggregates
may grow and shrink and their structure depends on the packing fraction of
the colloidal particles in the aggregates and
on the strength of the attraction among the colloidal particles.
In general the aggregates may consist of crystalline structures,
gels, glasses, but also loose fractal aggregates may occur.
In the present paper we shall address the possibility of aggregation
of colloids immersed in near-critical binary liquid mixtures
serving as their solvent.

Beysens and Est{\`e}ve \cite{Beysens-et:1985} have reported
aggregation phenomena for silica spheres immersed in a binary
liquid mixture of water and lutidine which exhibits a miscibility
gap with a lower critical point \cite{Cox-et:1956,Jayalakshmi-et:1994}.
Using light scattering these authors found that upon approaching the
bulk coexistence region of demixing from the one-phase region of the
binary liquid mixture at constant composition of the solvent, the
silica spheres coagulate reversibly at rather well defined temperatures.
Moving back the thermodynamic state deeply into the one-phase region
the coagulated aggregates dissolve again. These observations, 
i.e., the formation of colloidal aggregates which sediment,
have been confirmed later 
for silica and polystyrene particles immersed in the same solvent 
\cite{Gurfein-et:1989,gallagher:92,Broide-et:1993,narayanan:93,Kurnaz,Jayalakshmi-et:1997}
and have been reported also for other systems
\cite{narayanan:93,Jayalakshmi-et:1997,Kline-et:1994,Koehler-et:1997,
grull:97,Bonn-et:2009,Buzzaccaro-et:2010};
for reviews see Ref.~\onlinecite{Beysens-et:1999}. 
The experimental findings have been interpreted as (the precursors of) 
a phase transition in the ternary mixture 
\cite{Kline-et:1994,Jayalakshmi-et:1997,Koehler-et:1997} or as a
(non-equilibrium) coagulation, which can be caused
by various mechanisms.
Some aspects of these experimental observations will be
discussed in more detail in Sect.~\ref{sec:comp_exp}.

Generically, colloidal particles have a preference for one of the
two components of the binary solvent. At the surface of the colloid
this preference gives rise to a surface field conjugate to 
the order parameter at the surface and thus leads to an adsorption 
layer rich in this component. Upon approaching the critical point
along the two-phase coexistence curve, at the wetting transition the
adsorption layers on the surface of the spherical colloids become
thick (but stay finite) so that they can snap  if the colloids get
sufficiently close \cite{bbd}. This leads to attractive interactions
\cite{bbd} and can give rise to coagulation.

Close to the critical point CCFs provide an additional mechanism
for mutual attraction between the colloids. Experimentally it has been 
observed that (strong) coagulation occurs for compositions of the
{\it{}s}olvent for which the concentration $c_{a}$ of the
{\it{}a}dsorbed component preferred by the colloids is comparable but 
less than its value at the critical point, i.e., when
$c_a\lesssim c^{(s)}_{a,c}$.
This observation is in line with both explanations mentioned above because the 
wetting transition is indeed located in that region of the bulk phase diagram of 
the solvent and because the CCFs are stronger there \cite{colloids1b} 
than for $c_a\gtrsim c^{(s)}_{a,c}$.
Several theoretical approaches have been used in order to study this
aggregation phenomenon for colloids immersed in a solvent which is
close to its phase separation
\cite{Sluckin:1990,Netz:1996,Loewen:1995,Law:98,Okamoto-et:2011}.
Most of them, however, stay on a qualitative level of description,
pointing out the possibility of coagulation.

Here we present a {\it quantitative} analysis of the stability of 
monodisperse spherical colloidal particles immersed in a critical 
binary solvent. 
Our analysis is based on controlled approximations
which we have worked out in Ref.~\onlinecite{MMD_part1}.
Following Ref.~\onlinecite{MMD_part1} 
we adopt a description in terms of an effective one-component system.
Within this approach the presence of the solvent is taken into account 
via an effective pair potential of the colloids.
For the regular background interaction potential we consider a generic,
softly repulsive contribution. 
The CCFs depend on the configuration of all colloidal particles
and thus are in general non-additive.
We focus on dilute suspensions with low colloidal number density
$\rho$, for which the approximation of pairwise additive CCFs is valid.
The critical Casimir pair potential between two spherical particles 
immersed in a binary liquid mixture is described in terms
of a universal scaling function corresponding to the Ising universality
class with symmetry-breaking boundary conditions 
\cite{FSS,Diehl:1986,Krech:1990:0,Dbook,colloids1a,colloids1b,Gambassi:2009}.
We apply the Derjaguin approximation, within which the pair potential
between two spheres is expressed in terms of the scaling function of 
the CCF in a slab \cite{Derjaguin:1934,colloids1a,Gambassi-et:2009}. 
The Derjaguin approximation can be applied, if the surface-to-surface
distances $D$ between the objects on which the CCF acts, i.e., 
$D\lesssim\xi$, are small compared with the radius $R$ of the colloids.

In $d=3$ and for a vanishing bulk ordering field $\hb=0$, the 
corresponding scaling function is known from Monte Carlo simulations
\cite{vas,Hasenbusch,Hasenbusch-cross,vas-cross}, from field-theoretical
studies \cite{krech}, and from the extended de Gennes-Fisher local-functional
method \cite{upton,FdeG_loc_fun}. Corresponding experimental data 
\cite{Hertlein-et2008,Nellen-et:2009,Gambassi-et:2009,pershan} are
also available. The conjugate ordering field \hb{} of the order
parameter of the solvent corresponds to the chemical potential
difference of the two species forming the binary liquid mixture. The
reversible aggregation phenomenon of colloids is observed for thermodynamic
states of the solvent which correspond to a nonzero \hb{}.
In order to handle the dependence of the CCF on this ordering field within the Derjaguin
approximation, we use still another approximation for the film scaling function of the 
CCF in $d=3$ (see, c.f., \eref{eq:sfp_h_approx}).
The scaling functions resulting from this approximation are comparable
with the ones obtained using the extended de Gennes-Fisher local-functional
method \cite{tobepublished} and are in qualitative agreement with
corresponding curves provided in Ref.~\onlinecite{Buzzaccaro-et:2010}
which are based on a density functional approach.

Our presentation is organized as follows.
In Sect.~\ref{ssc:coll_int} we shortly summarize the effective 
one-component description which we use for our analysis.
For more details and a comprehensive discussion of the validity
and of the limitations of this description we refer to the preceding 
paper \cite{MMD_part1}. In Sect~\ref{ssc:shape} we discuss the 
possible functional forms of the effective interaction potential
employed in our study.
Section~\ref{ssc:res_stability} is concerned with aggregation
phenomena in colloidal suspensions and in
Sect.~\ref{ssc:res_structure} results for the radial
distribution function of the colloids are presented.
In Sect.~\ref{sec:comp_exp} we compare available experimental data
with our results. We summarize our study in Sect.~\ref{sec:summary}.

\section{Effective interactions \label{ssc:coll_int}}

We aim for a theoretical description of the behavior of monodisperse
colloids immersed in a near-critical solvent. To this end we consider 
a monodisperse, effective one-component system of colloids with radius $R$
for which the presence of the solvent enters via the following effective
pair potential $V\of{r=D+2R}/(k_BT)=\pot\of{r=D+2R}$: 
\begin{widetext}
\begin{equation}
\label{eq:coll_potential0}
   \pot\of{r}=
    \begin{cases}
      \infty, & D<0 \\
      \pot[rep]+\pot[c]^{(d=3)}=
      A\exp(-\kappa D)+(1/\Delta)\theta^{(d=3)}(\Theta,\Delta, \Sigma),  & D>0.
    \end{cases}
\end{equation}
\end{widetext}
In \eref{eq:coll_potential0} the background interaction
potential between the colloids, which is present also away from \tcb{},
consists of a hard core repulsion for center-to-center distances $r<2R$
and a soft, repulsive tail \pot[rep] for $D=r-2R>0$. The latter prevents
aggregation due to omnipresent effectively attractive dispersion forces.
In the case of electrostatic repulsion, the range $\kappa^{-1}$ of
the repulsion is associated with the Debye screening length
and its strength $A$ depends on the colloidal surface charge
density \cite{Barrat-et:2003,Russel-et:1989}. 
We do not consider an additional interaction which would account for
effectively attractive dispersion forces. Effectively, dispersion forces
can be switched off by using index-matched colloidal suspensions. 
As will be discussed in the next section,
the presence of attractive dispersion forces does not change the
conclusions of the present study.
The critical Casimir potential $\pot[c]^{(d=3)}$ has a scaling form
with $\theta^{(d=3)}$ as its universal scaling function.
$\theta^{(d)}$ depends on the spatial dimension $d$ and
on the scaling variables $\Theta=\sgn\of{t} D/\xi$, $\Delta=D/R$, and
$\Sigma=\sgn\of{\hb}\xi/\xi^{\of{h}}$.
The bulk correlation lengths 
$\xi\of{t\gtrless 0}=\xi_0^{\pm}\abs{t}^{-\nu}$ and
$\xi^{\of{h}} =\xi^{\of{h}}_0\abs{\hb}^{-\nu/\of{\beta\delta}}$ 
govern the exponential decay of the bulk solvent OP correlation function
for $t\to0^{\pm}$ at $\hb=0$ and for $\hb\to 0$ at $t=0$, respectively; 
one has $t=(\tcb-T)/\tcb$ for a lower critical point.
The amplitudes $\xi_0^{\pm,h}$ are non-universal but related via 
universal amplitude ratios; $\nu$, $\beta$, and $\delta$ are standard
bulk critical exponents \cite{Pelissetto-et:2002}.

For $\theta^{(d=3)}$ we use the Derjaguin approximation. For the configuration 
of two spheres it renders \cite{Derjaguin:1934,colloids1a,Gambassi-et:2009}
\begin{equation}
\label{eq:sfpot_derjaguin}
	\sfp[\of{d=3}]_{Derj}\of{\Delta,\Theta,\Sigma}= 
		\pi \int_1^{\infty}\of{x^{-2}-x^{-3}}
                \sff[\parallel]^{\of{d=3}}
                     \of{x\Theta,\Sigma} \text{ d}x; 
\end{equation}
$\sff[\parallel]^{ \of{d}}\of{y,\Sigma}$ 
is the universal scaling function of the CCF \ccf[,\parallel] per 
$k_B\tcb$ and per area $A$ for a slab of the thickness $L$:
$\ccf[,\parallel]/A=k_{B} \tcb L^{-d}
      \sff[\parallel]^{\of{d}}\of{y=\sgn\of{t}L/\xi,\Sigma}$. 
From \eref{eq:sfpot_derjaguin} it follows that the scaling function 
$\sfp[(d=3)]_{Derj}$ does not depend on $\Delta$, which therefore
enters into \eref{eq:coll_potential0} only as a prefactor.
In the cases in which $\Sigma=0$, i.e., for a vanishing 
bulk field \hb{} we use $\sff[\parallel]^{\of{3}}\of{y,\Sigma=0}$ 
as determined from Monte Carlo simulations
\cite{vas,Hasenbusch,Hasenbusch-cross,vas-cross}. 

The scaling function for nonzero $\Sigma$ is 
obtained from the following approximation (see Eq.~(24) in 
Ref.~\onlinecite{MMD_part1} and the subsequent discussion there):
\begin{equation}
\label{eq:sfp_h_approx}
	\sff[\parallel]^{\of{d}}\of{y,\Sigma}\simeq
		\sff[\parallel]^{\of{d}}\of{y,\Sigma=0}
		\frac {\sff[\parallel]^{\of{d=4}}\of{y,\Sigma}}
		      {\sff[\parallel]^{\of{d=4}}\of{y,\Sigma=0}}.
\end{equation}
Within this approximation, for $\hb\to 0$ the 
scaling function reduces exactly to
$\sff[\parallel]^{\of{d}}\of{\Sigma=0}$ (for all $d$) and for fixed 
values of $y$ its functional form is the one obtained from
mean-field theory (MFT, $d=4$). 
The necessary input for this approximation is the mean-field scaling
function of the CCF for the film geometry, which follows from the
field-theoretical approach within the framework of the Landau-Ginzburg theory.
For colloids strongly preferring one and the same of the two species of
the binary liquid mixture the so-called strong adsorption limit applies.
Accordingly, the appropriate boundary conditions are the ones breaking the
symmetry of the OP at the colloid surfaces.
We determine the CCF from the local OP profile \Op[MFT] within
MFT via the so-called stress-tensor \T{} \cite{Eisenriegler-et:1994}.
The CCF per area in a slab, which is confined along the $z$-direction, is given 
by the \of{z,z} component of the stress tensor,
$\ccf[,\parallel]^{\of{MFT}}/A=
    {k_B \tcb}\thermal{\T[z,z]\fd{\Op[MFT]}-\T[z,z]\fd{\Op[b,MFT]}}$ 
(see Eqs.~(20) and (25) in Ref.~\onlinecite{MMD_part1}).

Our results will be expressed in terms of the parameters
\begin{equation}
\begin{split}
  \label{eq:parametrization}
 	& s=\kappa R, \qquad a=A/s, \qquad \zeta=\sgn\of{t}\kappa \xi, \\
 	&\text{and} \qquad  x=\kappa D=\kappa r -2 s.
\end{split}
\end{equation}

Within this parametrization the functional form 
of \pot{} is determined by $a$, $\zeta$, and $\Sigma$, while $s$ tunes 
the overall strength of the potential without affecting its shape:
\begin{widetext}
\begin{equation}
\label{eq:coll_potential}
	\pot\of{r=D+2R}=
		\begin{cases}
		\infty & x<0 \\
		s \set{a \exp\of{-x}
                  +(1/x) \sfp[\of{d=3}]_{Derj}\of{x/\zeta,\Sigma}
                  } & x>0.
		\end{cases}
\end{equation}
\end{widetext}
The parameter $\zeta$ can be varied experimentally by changing the
temparature (via $\xi$) or by adding salt to the solution 
(via $\kappa$); it expresses the competition between the length scales
of electrostatic repulsion and of the 
attractive CCF. The strength of
repulsion is measured by $a$; it is usually kept constant. $\Sigma$ is
determined by the thermodynamic state of the solvent. The ranges of the 
values of the parameters entering into the effective potential 
(\eref{eq:coll_potential}) and of the scaling variable $\Sigma$, which
correspond to possible experimental realizations, are discussed in
detail in Ref.~\onlinecite{MMD_part1}. The divergence
$\sim x^{-1}$ in $\pot\of{D=x/\kappa \to 0}$ (see \eref{eq:coll_potential})
is unphysical \cite{MMD_part1}, because the scaling form used for $\pot[c]^{\of{d}}$
applies only for distances large compared with microscopic length
scales such as $\xi^{\pm}_0$. In order to
circumvent this unphysical behavior we shall apply a linear
extrapolation scheme for $\pot\of{D\to 0}$ whenever this is necessary.

\section{Shapes of the effective potential}
\label{ssc:shape}

\begin{figure}
	\includegraphics{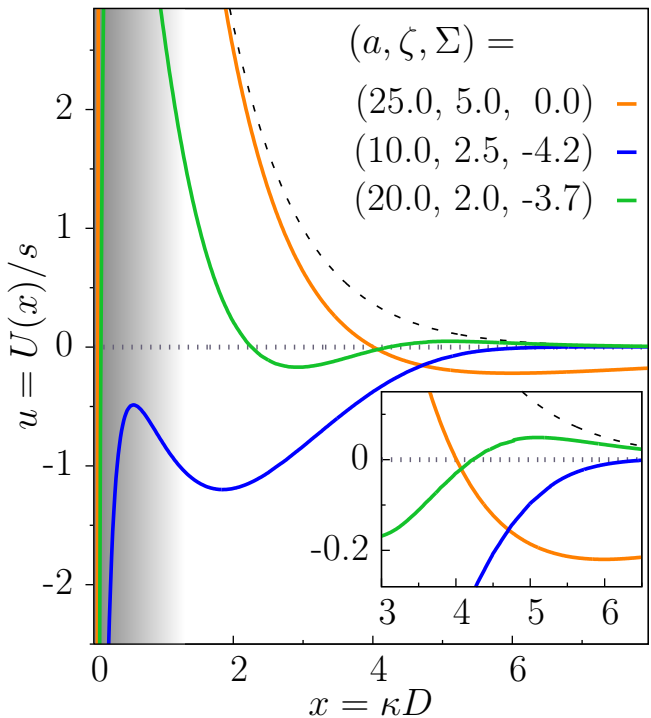} 
	\caption
	{Exemplary shapes of the effective pair potential given by \eref{eq:coll_potential}.
	 The dashed line represents the purely repulsive part of the potential for $a=20$.
	 The potentials considered here are physically
	 appropriate for $x=\kappa D\gtrsim 1$.
	 At smaller values of $x$ (grayish area) additional (effective) forces can occur.
	 In particular, the occurrence of a first minimum at contact, i.e., at $D=0$,
	 depends on microscopic details, which are not taken into
	 account by \eref{eq:coll_potential}.}
	\label{fig:coll_potential}
\end{figure}

By varying the thermodynamic state of the solvent the range and the strength of
the attractive CCFs are changed. This variable attraction combined with the
repulsive background contribution results in a broad variety of
possible shapes of the effective potential given by \eref{eq:coll_potential}.
Examples of various shapes of this effective potential, as obtained for
different values of the parameters $a$, $\zeta$, and $\Sigma$, are shown
in \fref{fig:coll_potential}, where we plot the reduced potential
$u=\pot/s$ versus the reduced distance $x=\kappa D$.
Depending on the values of the parameters, the CCFs compensate the repulsion
for all values of $D$ or only within certain regimes.
Compensation for all values of $D$ occurs if the strength $a$ of the repulsion is
not too strong.
This is illustrated by the blue curve in \fref{fig:coll_potential}
corresponding to the parameter values
$a=10$, $\zeta=2.5$, and $\Sigma=-4.2$.
We recall that negative values of $\Sigma$ correspond to solvent compositions
which are poor in the species preferred by the colloids.
For larger values of $a$, a secondary attractive minimum (i.e., $u<0$) can
occur (in addition to the primary global minimum for $D\to 0$, 
which will be discussed below), while for small
distances $D$ the potential maintains a repulsive part, as for the 
orange and green curves in \fref{fig:coll_potential}. Since $s=\kappa R$ 
can vary between ca. $10$ and $100$ 
a shallow minimum in $u$ can correspond to a deep potential well in $\pot$.
For suitably chosen parameters, such as for the green curve
($a=20$, $\zeta=2$, $\Sigma=-3.7$)
in \fref{fig:coll_potential}, the potential may be repulsive at large
distances $D\to \infty$ and may have an attractive secondary minimum.
Thus the CCFs provide a mechanism for realizing
short-ranged attractive and long-ranged repulsive interaction
pair potentials as studied for example in 
Refs.~\onlinecite{Archer-et:2007,Ciach:2008} in related contexts.
These studies have shown that such potentials give rise to complex
phase behaviors of colloidal suspensions. 

The issue of the contact value $\pot\of{D=0}$ and the possibility
of the occurrence of a primary minimum at small values of $D$ is 
subtle and, as pointed out in Sec.~\ref{ssc:coll_int}, cannot be
discussed in general in terms of the effective potential
\pot{} given by \eref{eq:coll_potential}. 
However, \pot{} captures the occurrence of
the potential barrier for $D\simeq \kappa^{-1}$ and
its reduction due to the attractive CCFs which is sufficient
for the focus of the present study, i.e.,
concerning the influence of the universal CCFs.
The short-ranged interactions acting at distances
$D\ll\kappa^{-1},\xi$ (with $\xi\gg \xi_0$) depend sensitively on
materials properties. But for the following reason they do
not affect our consideration of the stability of colloidal suspensions.
In the case of large $a$,
the repulsion is strong enough such that the CCFs cannot significantly 
reduce the associated potential barrier. 
Therefore within experimental observation times only
the pair-interaction potential \pot{} considered in
\eref{eq:coll_potential} is relevant for determining the phase behavior
and the coagulation process. The influence of attractive short-ranged
forces could be observed only for very long waiting-times because the 
time scale for passing the barrier is $\sim\exp\of{\pot{}}$ (see 
Subsect.~\ref{ssc:res_stability} below).
On the other hand in the case of small $a$ the CCFs fully compensate
the repulsion. The estimates for the values 
of the parameters for which the repulsive barrier disappears and
thus rapid coagulation sets in can be obtained from the effective
potential \pot{} given in \eref{eq:coll_potential}.
However, in order to be able to determine the detailed process of
coagulation a truly microscopic description is needed. This is beyond
the scope of the present study devoted to {\it universal} features.
As long as no additional microscopic forces compete with the considered
effective potential, \eref{eq:coll_potential} provides a 
description which is sufficient for the present purpose.

\section{Stability \label{ssc:res_stability}}

The stability of colloidal suspensions and the aggregation of colloids are related to
kinetic processes (see Ref.~\onlinecite{Russel-et:1989} and references therein).
They are based on the diffusion of single particles in the presence
of other particles of the same kind, interacting with them via
interaction potentials which contain both attractive and repulsive contributions.
For interacting particles, which irreversibly stick together once their surfaces
touch each other, Fuchs introduced, using Smulochowski's theory,
the stability ratio \w{} \cite{Fuchs:1934},
\begin{equation}
  \label{eq:stabilityratio}
	\w = 2R \int_{2R}^{\infty}\frac{\exp\set{\pot\of{r}}}{r^2} \text{ d}r.
\end{equation}
For hard spheres $\w=1$, while for $\w>1$ ($\w<1$) the repulsive (attractive) 
part of the interaction potential \pot{} dominates. If there is a potential 
barrier, i.e., $\pot\of{r}\gg 1$ for a certain range of distances $r>2R$, 
which leads to $\w>1$, on intermediate time scales the suspension will 
equilibrate into a (meta)stable homogeneous state. Only on very large time 
scales cluster formation sets in slowly. It turns out that \w{} is proportional 
to the ratio between the characteristic times for diffusion of a single particle, 
$t_{diff}$, and for the formation of a pair of particles, $t_{pair}$ \cite{Russel-et:1989}:
\begin{equation}
\label{eq:stability_timescales}
   t_{diff}/t_{pair}=3 \eta/\w,
\end{equation}
with the packing fraction $\eta=\of{4\pi/3}R^{3}\rho$; $\rho$ is the 
number density of the colloidal particles.

\begin{figure}
	\includegraphics{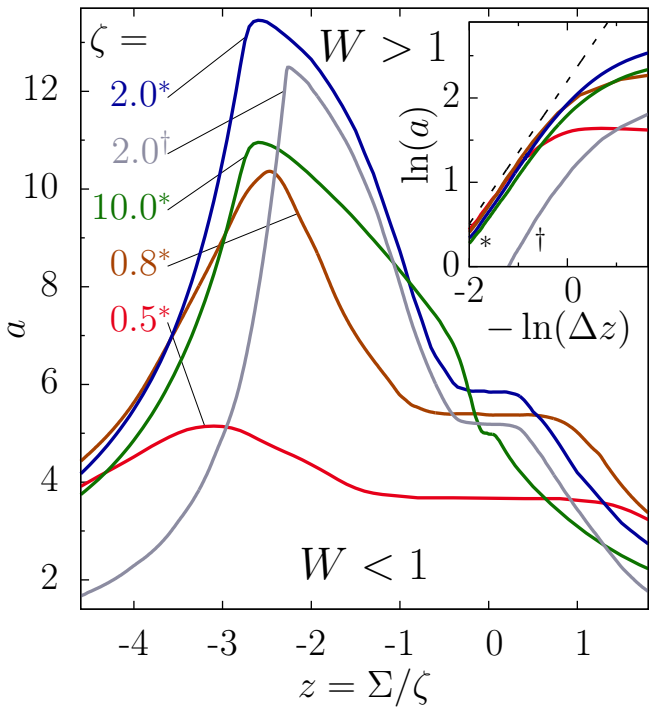} 
	\caption{Contour lines of constant stability ratio $\w=1$
		(\eref{eq:stabilityratio}) for the effective pair potential
		given in \eref{eq:coll_potential} as function of 
		$a$ and $\Sigma$ for various (fixed) values of $\zeta$. 
		The star and the dagger superscripts
		correspond to $s=10$ and $s=50$, respectively.
		For values $\w<1$ (the region below the corresponding 
		contour line) the colloids coagulate rapidly.
		For a binary solvent with compositions 
		which are smaller, i.e., slightly poorer
		in the component preferred by the colloids, than the 
		critical composition, i.e., for
		$\Sigma \lesssim 0$, this region ($\w<1$) broadens 
		towards larger values of the parameter $a$ of the 
		repulsion strength. The logarithmic plot in the inset 
		shows, that for rather large negative values of $\Sigma$
		the contour lines decay according to a power law as
		function of $\Delta z=z_{m}-z$.
		The slope of the dashed line is $0.84$;
		$z_{m}=-2.6$ approximates the position of the maxima of 
		the contour lines.
		 }
	\label{fig:stability_diagram}
\end{figure}
We have calculated the stability ratio $\w=\w\of{a,s,\zeta,\Sigma}$ 
according to \eref{eq:stabilityratio} by using \eref{eq:coll_potential}
within the one-phase region. Since the stability ratio \w{} depends 
 exponentially on the potential, in the calculation 
of \w{} we can afford to omit the use of a cut-off at small distances $D$.
The main contribution to \w{} stems from the maximum of the potential. 
The use of a cut-off would only slightly change the value of \w{} but
it would require an additional specification which is not necessary
for this analysis.

In \fref{fig:stability_diagram} the contour lines $\w=1$ are plotted in
the space spanned by the parameters $a$ and $z=\Sigma/\zeta$
for various values of $\zeta$. Changing $\Sigma$ for fixed $\zeta$ and $s$
corresponds to moving along an isotherm in the thermodynamic space
of the system. In order to have the contour lines on the same scale, it
is useful to introduce the reduced variable $\Sigma/\zeta$ as abscissa.
For $\w=1$ the time scales of diffusion and pairing of two particles
are comparable (\eref{eq:stability_timescales}).
For a given value of $\zeta$, in the region $\w>1$ above the corresponding
contour line the suspension initially equilibrates 
into a (metastable) fluid state. 
If the potential has its global minimum at $D=0$, e.g., due to 
van-der-Waals forces, aggregates may form eventually at long times;
these time scales depend on the value of \w. 
On the other hand, for $\w{}<1$ rapid coagulation sets in.

The overall characteristics of the shape of contour lines 
$\w=1$ remains the same for different (but fixed) values of
$\zeta$ and $s$ (see \fref{fig:stability_diagram}).
Upon decreasing $\Sigma$ towards slightly negative values
(i.e., for a binary solvent upon decreasing the amount of the
component preferred by the colloids relative to the critical composition)
the colloidal stability remarkably decreases, i.e., the contour line
$\w=1$ is shifted to larger values of the parameter $a$ of the
repulsion strength. For even more negative values of $\Sigma$,
the system is driven far away from criticality, so that the attraction
due to the CCFs weakens and stability is favored again.
The position $z_{max}$ of the maximum of the corresponding line
$a=a\of{z=\Sigma/\zeta}$
(given by $\w\of{a,s,\zeta,\Sigma=z\zeta}=1$ for fixed $s$ and fixed $\zeta$)
does hardly vary with $\zeta$.
However the amplitude of $a\of{z}$ and in particular the maximum value
$a_{max}=a\of{z_{max}}$ exhibit a significantly
nonmonotonous dependence on $\zeta\sim\abs{t}^{-\nu}$.
The value of $a_{max}$ is largest for $\zeta\simeq 2$.
This behavior reflects the fact that for \of{+,+} boundary conditions 
(i.e., equal surface fields on both surfaces) as considered 
here the CCFs are not strongest at \tcb{} and
$\hb=0$, but slightly within the one-phase region and
for small negative values of $\hb$.

The contour lines corresponding to a fixed value of $s$ and
various values of $\zeta$ merge to a single curve for large
negative and large positive values of $\Sigma/\zeta$.
 This behavior can be explained by the fact that
for large values of $\abs{\Sigma}$ the scaling functions
$\sfp[\of{d=3}]_{Derj}\of{\Theta=D/\xi,\Sigma}$
plotted as a function of $\Theta\Sigma$ for various
values of $\Sigma$ fall onto each other \cite{tobepublished}.
Interestingly, within the plotted range of
large negative values of $z=\Sigma/\zeta$ the (common)
decay of the contour lines can be described by a power law
$\sim\of{z_{m}-z}^{-p}={\Delta z}^{-p}$ (see the double logarithmic 
plot in the inset of \fref{fig:stability_diagram}, where $z_{m}=-2.6$ 
approximates the various $z_{max}\of{\zeta}$ and $p \simeq 0.84$).
In the limit of strong bulk fields, i.e.,
$\abs{\hb}\to\infty$, the CCFs are vanishingly small.
In this limit, $\w=1$ for HS without a soft repulsive contribution 
and therefore $\of{a\of{\abs{z}\to\infty}}_{\w=1}\to 0$.
The leveling off of the contour lines $\w=1$ close to $\Sigma=0$ is
a consequence of the choice of the abscissa
$\sim {\hb}^{\nu/\of{\delta\beta}}$.
As a function of \hb{} the contour lines exhibit no
leveling off around $z=0$.

The potential under consideration (within the Derjaguin approximation) is
proportional to the parameter $s$. Thus with increasing (decreasing)
values of $s$ the region $\w<1$ of aggregation shrinks (widens),
because the repulsive barrier in $\pot$ becomes larger (smaller).
However, since the repulsive barrier contributes exponentially to
\w{}, the boundary of the region $\w<1$ is mostly determined by the
sheer occurrence of the repulsive barrier (which, however, is
independent of $s$) and therefore the dependence of the contour line
$\w=1$ on $s$ is weak. This is seen in \fref{fig:stability_diagram}
where for $\zeta=2$ we compare the contour lines $\w=1$ for $s=10$
and $s=50$ (the blue and gray curves, respectively).
Note, that this weak dependence of the stability region on the parameter
$s=\kappa R$ is in contrast to the strong dependence on $s$
of the two-phase region of the effective colloidal system, as discussed
in Ref.~\onlinecite{MMD_part1} (see especially Fig.~3 therein).
For the two-phase region not only the existence but also the actual
strength of the attraction is important, and therefore it is sensitive to
the value of $s$, because $\pot\sim s$.

The contour lines in \fref{fig:stability_diagram} of constant stability
ratio $W=1$ can be compared with the contour
lines in Fig.~3 in Ref.~\onlinecite{MMD_part1} of constant critical temperature 
$T_c^{(eff)}$ (or $\zeta_c^{(eff)}$) delimitating 
the "liquid``-``gas" coexistence of the effective one-component 
colloidal system. 
In both figures the contour lines are plotted in the plane spanned by the 
parameter $a$, measuring the strength of the repulsion of the effective 
pair potential, and the scaling variable 
$\Sigma$ (or $\Sigma$ rescaled by the fixed value of $\zeta$). 
The gross features of these contour lines are similar 
but the values of $a$ are rather different. 
(Note the different scales for $a$ in these two figures.) 
For example, for the same value of $s=\kappa R=10$, at the temperature corresponding
to $\zeta=2$ strong coagulation (i.e., $W<1$) occurs for $a\lesssim 13$, whereas at the
comparable temperature corresponding to $\zeta=3$ the colloidal
suspension  starts to phase separate already for a much stronger repulsion,
i.e., for $a \lesssim 300$ (see the blue curves in Fig.~3 in Ref.~\onlinecite{MMD_part1}). 
We note, however, that phase separation occurs at relatively
high packing fractions $\eta \gtrsim 0.06$ of the colloidal particles (see also
Subsect.~III~B in Ref.~\onlinecite{MMD_part1}) whereas aggregation occurs
also for very small packing fractions of the colloidal particles.

\begin{figure}
	\includegraphics{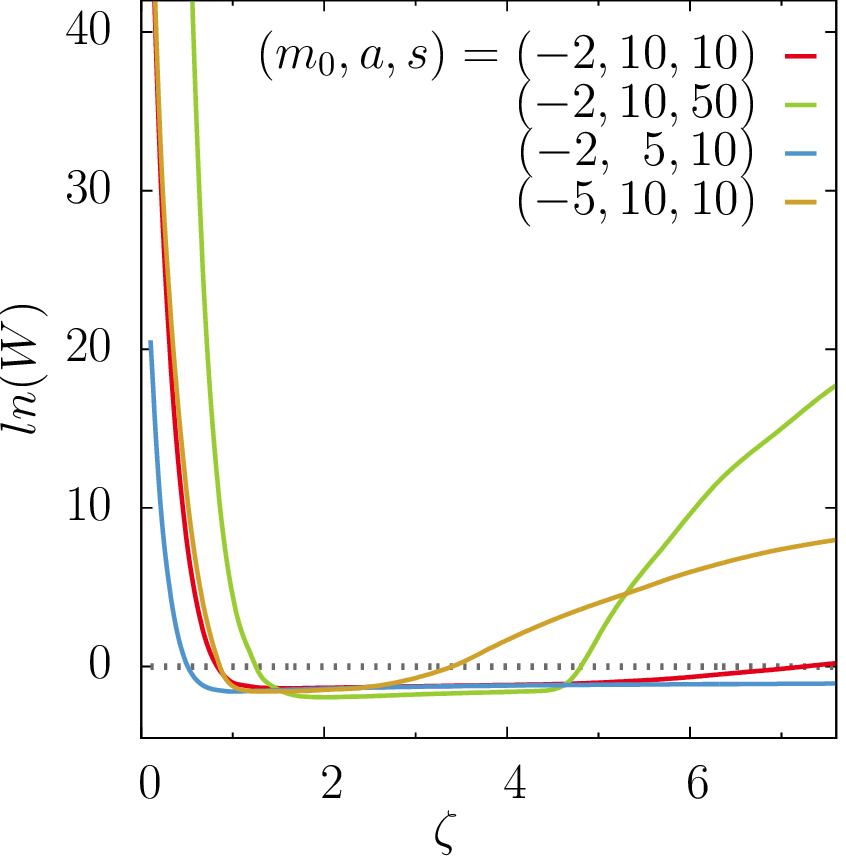}
\caption{
	Semi-logarithmic plot of the stability ratio 
	$\w$ (\eref{eq:stabilityratio}) for the typical
	experimental situation of varying 
	the reduced temperature $\abs{t}\sim \zeta^{-1/\nu}$
	at fixed composition $\of{c_a-c^{(s)}_{a,c}} \sim {m_0}^{-1}$
	(\eref{eq:op_amplitude_modified}).
	The steep decrease of $\ln\of{\w}$ towards negative values
	upon increasing $\zeta$, i.e., for $t\to 0$, suggests that 
	coagulation sets in within a narrow temperature interval.
	The increase of $\w$ for large values of $\zeta$ occurs for
	strongly repulsive contributions to
	the effective potential
	($a=10$, \eref{eq:coll_potential});
	$s=\kappa R$, where $\kappa^{-1}$ is the range of the repulsive
	``background'' potential and $R$ is the 
	radius of the colloids.
	}
	\label{fig:stab_temperature}
\end{figure}
The experimentally relevant case of the variation of the stability ratio 
\w{} with temperature at constant OP is shown in a semi-logarithmic plot 
in \fref{fig:stab_temperature} in terms of $\zeta\sim\abs{t}^{-\nu}$ and 
the parameter $m_0$ which is related to the OP; for $t>0$
it is given by (see Eq.~(27) in Ref.~\onlinecite{MMD_part1}) 
\begin{equation}
\label{eq:op_amplitude_modified}
m_0=\sgn\of{\phi}\of{\zeta_0^{+}}^{1/\nu}\abs{\mathcal{B}/\phi}^{1/\beta}
\end{equation}
and $\zeta_0^{+}=\kappa \xi_0^{+}$.

\Op{} and \hb{} are related by the equation of state,
$\hb=\mathcal{D}\sgn(\Op)\abs{\Op}^{\delta}
  \sfeos\of{t \abs{\mathcal{B}/\Op}^{1/\beta} }$, where
$\sfeos\of{\hat{X}}=1+\hat{X}$ (in lowest
order in $\hat{X}$) is a universal (bulk) scaling function
(see Appendix~A in Ref.~\onlinecite{MMD_part1}). 
$\mathcal{B}$ and $\mathcal{D}$ are non-universal amplitudes which
depend on the definition of $\phi$ such that on the coexistence curve
the bulk OP follows
$\Op[b]\of{t\to0^{-},\hb=0}=\mathcal{B}\abs{t}^{\beta}$ and that on 
the critical isotherm
$\Op[b]\of{t=0,\hb\to 0}=\sgn\of{\hb}\abs{\hb/\mathcal{D}}^{1/\delta}$. 
For example, in binary liquid mixtures the order parameter \Op{} is
proportional to the deviation of the concentration
$c_a$ of the component $a$ from its critical value $c^{(s)}_{a,c}$,
$\Op= {\cal A}(c_a-c^{(s)}_{a,c})$ (note that $\mathcal B$ is
proportional to ${\cal A}$), which can be easily controlled by changing
the mass or the volume fraction of one of the components of the mixture. 

The steep decrease of \w{} upon increasing
$\zeta\lesssim1$ indicates that coagulation is expected to set in 
above the critical temperature (i.e., $\w$ crossing the value 1 
from above) within a rather narrow temperature interval.
For the plots in \fref{fig:stab_temperature} we have chosen
parameter values which allow one to focus on the region of
the onset of coagulation. As one can infer from the stability diagram
(\fref{fig:stability_diagram})
$a=10$ corresponds to a relatively strong repulsion in the sense that 
for this value the range of values of $\Sigma$ for which
$\w<1$ holds and thus coagulation occurs is rather narrow.
This implies that one finds $\w<1$ only for intermediate values of
$\zeta$. Such a case is represented by the red curve in
\fref{fig:stab_temperature} which corresponds to $m_0=-2$, $a=10$, and $s=10$.
We find a clear increase of $\w$ for $\zeta\gtrsim 4.5$;
for larger $\zeta$, $\w$ becomes even larger than $1$.
The effect of varying $m_0$, $a$, or $s$ relative to the red curve
can be inferred from the brown, blue, and green curves, respectively.
Since the height of the potential barrier is proportional to $s$,
stability is significantly enhanced for larger values of $s$
(see the green curve \of{s=50} in \fref{fig:stab_temperature}).
The value of $\zeta$, for which $\ln\of{W}$ turns from positive to negative
(negative to positive) is larger (smaller) than the corresponding one for $s=10$.
For the composition corresponding to $m_0=-5$, which is closer to the
critical composition (see \eref{eq:op_amplitude_modified}), the region
of values of $\zeta$ for which $\w<1$ is smaller than for $m_0=-2$
(see the brown curve in \fref{fig:stab_temperature}).
Upon decreasing $a$, the value of $\zeta$ for which $\ln\of{W}$ changes 
from being positive to being negative becomes smaller (see the blue 
curve \of{a=5}). For such small values of $a$, $\ln\of{\w}$ remains 
negative upon increasing $\zeta$ (also for values of the parameters
$m_0$ and $s$ not shown in \fref{fig:stab_temperature}). 

In summary, along the thermodynamic paths considered
(and typically realized in experiments)
the range of the attraction due to the CCFs grows steadily with
increasing $\zeta$, but the amplitude of the CCFs
is a nonmonotonous function of $\zeta$. The CCF attains its maximal
strength for an intermediate value of
$\zeta$. Therefore, upon increasing $\zeta$
the repulsive barrier in the effective pair potential
is at first reduced; for sufficiently small values of $a$ it can
even disappear altogether.
For even larger values of $\zeta$, the potential barrier 
may grow again (or emerge) because the amplitude of the CCF 
decreases again. This latter barrier is located at smaller values of $D$
than the one which is present for small values of $\zeta$.
The nonmonotonous dependence of the maximal strength of
the CCF on temperature results in a nonmonotonous
behavior of $\w=\w\of{\zeta}$.
Although, as stated above, a repulsive barrier may
emerge again, it is likely that aggregates, which have
formed for intermediate values of $\zeta$, will not break up.
This might be due to either specific microscopic interactions or due
to the secondary minimum in the effective pair potential,
which is still quite deep. Also in the case
in which for all values of $\zeta$ a repulsive barrier
remains, coagulation can appear, due to a deep secondary minimum.
This case will be discussed
in Sect.~\ref{sec:comp_exp}.

\section{Structure \label{ssc:res_structure}}
The bulk structure of a colloidal suspension is characterized by 
its radial distribution function $g\of{r}$ which can be interpreted as 
the probability to find a colloidal particle a distance $r$ apart from 
another particle fixed at the origin. The total correlation function 
(TCF) $h\of{r}$ is related to the former one according to 
$h\of{r}=g\of{r}-1$. The Ornstein-Zernicke equation expresses $h\of{r}$ 
in terms of the direct correlation function (DCF) $c\of{r}$ and the 
number density $\rho$ \cite{Ornstein-et:1914,Hansen-et:1976},
\begin{equation}
  \label{eq:oz}
  h\of{r}= c\of{r} + \rho \int 
	    h\of{r'} c\of{\abs{\x-\x'}}\text{ d}^3\x',
\end{equation}
or equivalently in Fourier space
\begin{equation}
  \label{eq:oz_fourier}
  \ft{h}\of{q}= \ft{c}\of{q} /\of{1-\rho \ft{c}\of{q}},
\end{equation}
with
$\ft{h}\of{\q}=\int e^{i\q\x}h\of{\x}\text{ d}^{d}\x$,
analogously for $\ft{c}\of{\q}$, and $q=\abs{\q}$.
In order to determine the two unknown functions $c\of{r}$ and $h\of{r}$ 
a second equation is required. This is the so-called closure equation 
which in its most general form is given by
\begin{equation}
	\label{eq:closure}
	h\of{r}+1=\exp \set{ -\pot\of{r} +h\of{r}-c\of{r} +b\of{r} }
\end{equation}
where $b\of{r}$ is the so-called bridge function, which in general is 
not known. The most common approximations for the bridge function are 
the so-called Percus-Yevick approximation (PY),
\begin{equation}
  \label{eq:py_bridgefct}
  b_{PY}\of{r}= \ln\fd{h\of{r}-c\of{r}+1}-h\of{r}+c\of{r},
\end{equation}
and the so-called hypernetted-chain approximation (HNC), 
\begin{equation}
  \label{eq:hnc_bridgefct}
  b_{HNC}\of{r}= 0.
\end{equation}
For practical reasons, in order to handle the hard core 
$\pot\of{r<2R}=\infty$, it is useful to introduce the function
\begin{equation}
\label{eq:k}
k\of{r}=h\of{r}-c\of{r}.
\end{equation}
In terms of $k\of{r}$ the PY closure (\eref{eq:py_bridgefct}) can be written as
\begin{equation}
\label{eq:py_k}
	{c}_{PY}\of{r}= \left[\exp\set{-\pot\of{r}}-1\right]\of{k\of{r}+1}
\end{equation}
and the HNC closure (\eref{eq:hnc_bridgefct}) as
\begin{equation}
\label{eq:hnc_k}
	c_{HNC}\of{r}= \exp\set{-\pot\of{r} +k\of{r}}-k\of{r}-1 .
\end{equation}

The correlation functions can be calculated 
iteratively \cite{Hansen-et:1976,Gillian:1979}.
For a given approximate $c_{i}\of{r}$ one calculates
the TCF $h_i\of{r}$ according to \eref{eq:oz_fourier}.
By using a closure, i.e., by choosing a bridge function,
\eref{eq:closure} renders a DCF $c_{i+1}$ which typically
differs from $c_{i}$. This procedure is continued until satisfactory 
convergence is achieved. The initial guess $c_{i=1}\of{r}$ is 
guided by the shape of the direct interaction potential.

The applicability and reliability of this integral equation approach (IEA) 
is discussed in detail in Ref.~\onlinecite{Caccamo:1996}. 
For comparison with Monte Carlo simulations in the case of a pair potential 
with attractive and repulsive 
parts see, e.g., Ref.~\onlinecite{Archer-et:2007}, which discusses particles 
interacting with a pair potential containing attractive and repulsive 
Yukawa-like contributions $\epsilon_{i} e^{-\kappa_{i} r}/r$. 
The IEA is capable to reveal the rich phase behaviour of such systems.

The structure factor
\begin{equation}
\label{eq:structurefactor}
    S\of{q}=1+ \rho \ft{h}\of{q}= 1/\of{1-\rho \ft{c}\of{q}} 
\end{equation}
can be determined by scattering experiments. From sum rules it follows that
the isothermal compressibility $\chi_T$ is given by \cite{Hansen-et:1976}
\begin{equation}
\label{eq:compressibility}
    \lim\limits_{q\to 0}S\of{q}= \rho k_B T \chi_T.
\end{equation}
The system becomes unstable for $\chi_T\to \infty$, corresponding
to the critical point and, within MFT, to the spinodals in the 
phase diagram (see, e.g., Ref.~\onlinecite{Binder-et:1978}).

\begin{figure}
	\includegraphics{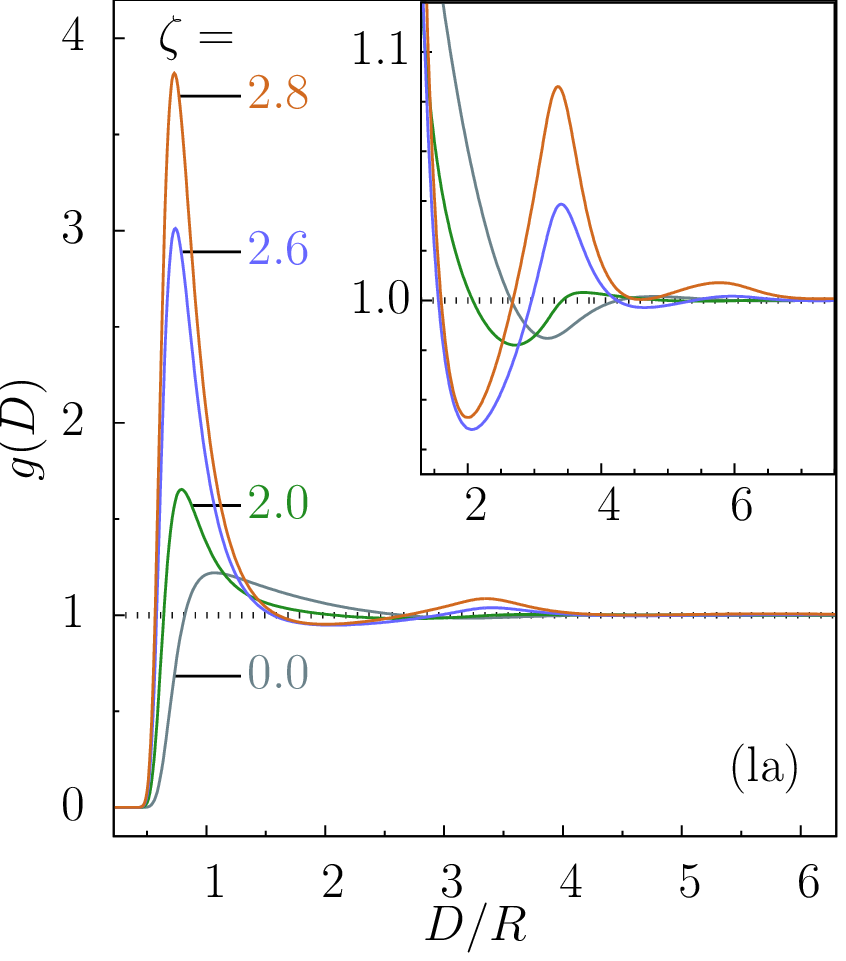} 
	\caption{The radial distribution function $g\of{r}$ as obtained 
	 from the HNC closure (\eref{eq:hnc_bridgefct}) for four values
	 of $\zeta\sim\abs{t}^{-\nu}$ and for a volume fraction
	 $\eta=0.04$ of the colloids.
	 The parameters of the corresponding effective potential
	 (attractive at large distances (la), \eref{eq:coll_potential}) 
	 are $a=100$ for the strength of the repulsion, $m_0=-100$
	 for the reduced composition (\eref{eq:op_amplitude_modified}), 
	 and $s=\kappa R=10$; $\kappa^{-1}$ is the range of the soft
	 repulsion, $R$ is the radius of the colloids, and $D=r-2R$ is the
	 surface-to-surface distance.
	 The sequence of curves corresponds to the generic isochoral
	 thermodynamic path realized in experiments.
	 For these values of the parameters,
	 upon decreasing the reduced temperature $t$, i.e.,
	 increasing $\zeta$, a single attractive minimum develops in the
	 effective potential.
	 This increasing attraction due to the CCF leads to the enhancement of
	 $g\of{r}$ close to the surface of the colloid.
	 This peak suggests that the formation of colloid dimers is favored.
	 The inset provides an enlarged view of the vicinity
	 of the second maximum of $g(r)$.
	}
	\label{fig:structure_simpleMin}
\end{figure}
\begin{figure}
	\includegraphics{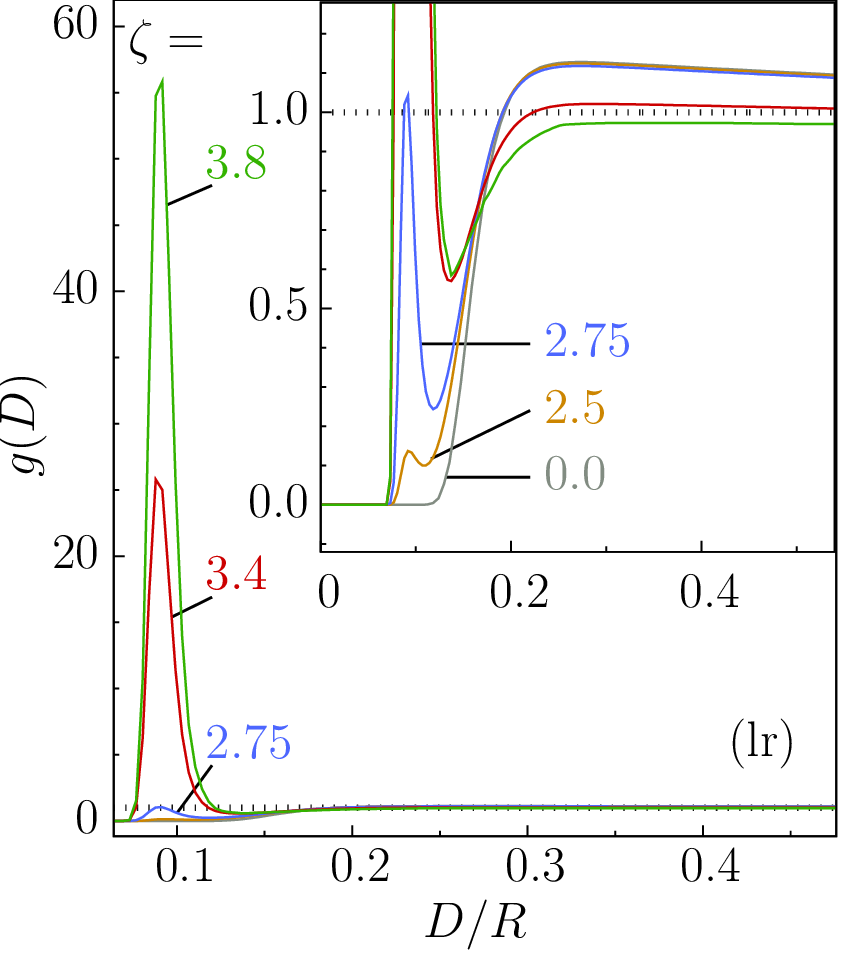}
	\caption{The radial distribution function $g\of{r}$ as obtained
	 from the HNC closure (\eref{eq:hnc_bridgefct}) for various values
	 of $\zeta\sim\abs{t}^{-\nu}$ and for a
	 volume fraction $\eta=0.04$ of the colloids.
	 The parameters of the corresponding effective potential
	 (repulsive at large distances (lr), \eref{eq:coll_potential}) 
	 are $a=40$ for the strength of the repulsion, $m_0=-3$ for
	 the reduced composition (\eref{eq:op_amplitude_modified}), 
	 and $s=\kappa R=50$; $\kappa^{-1}$ is the range of the soft repulsion,
	 $R$ is the radius of the colloids,
	 and $D=r-2R$ is the surface-to-surface distance.
	 For these values of the parameters, upon decreasing the
	 reduced temperature $t$, i.e., increasing $\zeta$,
	 an attractive minimum develops in the effective potential at
	 intermediate values of $D$ while repulsion
	 remains at small and large values of $D$.
	 Due to this attraction a peak develops upon decreasing $t$.
	 For a comparable potential well depth this peak is higher
	 and narrower than in the case without a long-ranged repulsion
	 (compare \fref{fig:structure_simpleMin}).
	}
	\label{fig:structure_sevExtrema}
\end{figure}

The diversity of possible shapes of the effective pair potential
given in \eref{eq:coll_potential} (see \fref{fig:coll_potential})
leads to a variety of radial distribution functions $g\of{r}$
(see Figs.~\ref{fig:structure_simpleMin} and \ref{fig:structure_sevExtrema}
in which the results for the HNC closure are shown;
the results obtained using the Percus-Yevick closure are almost
the same, differing mostly in that, for the same
value of $\zeta$, the amplitude of the resulting
$g\of{r}$ is slightly smaller than the one within the HNC
closure). 
For temperatures far away from the critical temperature of the solvent
(i.e., for $\zeta=\kappa\xi \ll 1 $), the colloids are
effectively hard spheres with an effective
diameter $\sigma>2R$ due to the soft repulsive background 
contribution \pot[rep]. Accordingly, for such values of $\zeta$, 
$g\of{r}$ has the corresponding characteristics of a fluid of 
hard spheres, such as the rather broad first peak for small values 
of $D$. Due to the emerging attractive CCFs, for increasing
$\zeta\sim \abs{t}^{-\nu}$, the radial distribution function
$g(r)$ is enhanced close to the surfaces of the colloids.
This implies an enhanced short-ranged order.

In the case that the effective potential exhibits a repulsive barrier 
at small values of $D$ and is {\it a}ttractive throughout {\it l}arge
distances (denoted by (la) in the following), we find that upon
increasing $\zeta$ the positions of the extrema of $g\of{r}$
move towards smaller values of $D$ (especially for intermediate 
values of $\zeta$) and the value of $g\of{r}$ at the maxima (minima) 
increases (decreases); the latter occurs mostly for larger values of 
$\zeta$ (see \fref{fig:structure_simpleMin}).

In contrast, in the case of effective potentials
which are {\it r}epulsive at {\it l}arge distances (denoted
by (lr) in the following), a new first peak at small
values of $D$ emerges and increases with increasing $\zeta$ while the
former first peak decreases and finally disappears
(see the brown curve for $\zeta=2.5$ in
\fref{fig:structure_sevExtrema} showing the emergence of a new peak
at $D/R\simeq{0.1}$; the broad former first peak is located at larger 
distances $D/R \simeq 0.25$).
A similar trend is seen for the other extrema at larger $D$.
However, since in that range of $D$ the amplitude of $g\of{r}$ is
already quite small, the actual changes of the shape of $g\of{r}$ 
cannot be described in such simple terms.

In homogeneous, i.e., not phase separated systems, the 
(newly emerged or shifted) first peak in $g\of{r}$ can become
significantly larger in the (lr) case than in the (la) case (see
Figs.~\ref{fig:structure_sevExtrema} and \ref{fig:structure_simpleMin},
respectively, and note the different scales).
The reason for this difference can be understood as follows.
To a large extent the height of the first maximum in $g\of{r}$
depends on the depth of the attraction well of the effective potential. 
Indeed, for comparable potential well depths the first peak in $g\of{r}$
is only slightly larger in the (lr) case than in the (la) case.
However, for equally deep potential wells the overall attraction is 
stronger in the (la) case due to the attraction throughout large distances
(which gives rise to a larger compressibility $\chi_T$ which
is proportional to the integral of $g\of{r}-1$). Accordingly, the 
systems interacting via (la) effective potentials can phase separate for
shallower potential wells. Systems with potential well depths corresponding 
to large first peaks in $g(r)$ (as the ones shown for the (lr) potentials in 
Fig.~\ref{fig:structure_sevExtrema}) and with a (la) character would already 
be phase separated. In this sense, the repulsion occurring at larger values 
of $D$ enhances the stability (w.r.t. phase separation) even of colloidal 
suspensions exhibiting large peaks in $g(r)$ in homogeneous systems.
In order to illustrate this point, we note that the compressibility 
corresponding to the blue curve ($\zeta=2.6$) in 
\fref{fig:structure_simpleMin} turns out to be approximately the same
as the one corresponding to the green curve ($\zeta=3.8$) in 
\fref{fig:structure_sevExtrema}, although for the latter 
system the peak in $g(r)$ is nearly $20$ times larger.

\section{Comparison with experiments} \label{sec:comp_exp}

For dilute suspensions, the second virial coefficient \cite{Hansen-et:1976} 
\bsec{} provides information about the strength
of the radially symmetric attraction between spherical particles:
\begin{equation}
\label{eq:secvirial}
   \bsec=2\pi \int_{0}^{\infty}\of{1-\exp\set{-U\of{r}}}r^{2} \text{ d}r.
\end{equation}
Beyond the ideal gas contribution it determines the leading non-trivial
term in the expansion of the pressure
$p\of{\rho}/\of{k_BT\rho}=1+\bsec{}\rho+\ldots$ in terms of powers of
the number density $\rho$.

\begin{figure}
	\includegraphics{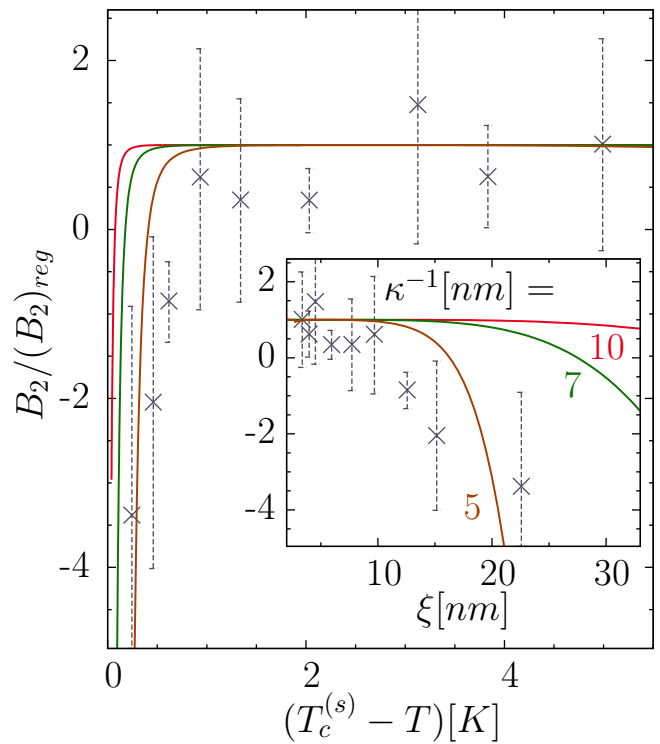}
	\caption{The value of the second virial coefficient $\bsec$
	(\eref{eq:secvirial}) normalized by its value $\of{\bsec}_{reg}$
	far away from \tcb{} as function of the temperature 
	deviation from the critical temperature \tcb{} of the solvent 
	and at its critical composition ($\Sigma=0$). 
	The same data in the inset but as function 
	of the correlation length $\xi=\xi_0\abs{1-T/\tcb}^{-\nu}$ of the
	solvent. The experimental values (symbols) for latex particles
	immersed in the critical water-lutidine mixture
	are taken from Fig.~4 of Ref.~\onlinecite{Kurnaz}(b).
	The lines correspond to the values obtained from the effective
	pair potential given in \eref{eq:coll_potential}. The values of
	$a$ and $s$ were obtained by using
	\eref{eq:A_electric} for $A$ and the values
	$\Upsilon=5.7\mu C/cm^2$ and $2R=555nm$ 
	as given in Ref.~\onlinecite{Kurnaz}(b). We have considered various
	values of $\kappa$ within the range reported in Ref.~\onlinecite{Kurnaz}(b).
	We have adopted the values  $\xi_0=0.25nm$ and 
	$\tcb=310 K$. For the critical mass fraction $\m[L,c]=0.287$ of lutidine {\cite{Kurnaz}$^{\text{(b)}}$} 
	one obtains \cite{note_permittivity} $\epsilon=22.3$.
	For further details see the main text.}
	\label{fig:b2_comparison}
\end{figure}
In Ref.~\onlinecite{Kurnaz} the second virial coefficient $\bsec$ has been
determined for latex particles immersed in a
mixture of 2,6-lutidine and water by using light scattering. In
\fref{fig:b2_comparison}, for the critical composition ($\Sigma=0$) we
compare the variation of $\bsec$ upon 
approaching the critical temperature \tcb{} 
for the effective pair potential in \eref{eq:coll_potential} with the
experimental data given in Fig.~4 of Ref.~\onlinecite{Kurnaz}(b). 
In order to be able to compare these data with $\pot\of{r}$ given in
\eref{eq:coll_potential0} we have to specify the amplitude $A$ of the 
repulsive contribution $\pot[rep]\of{r}=A\exp\of{-\kappa D}$. This amplitude
depends on the surface charge density $\Upsilon$ of the colloids \cite{Russel-et:1989}: 
\begin{equation}
\label{eq:A_electric}
	A= \of{\epsilon \epsilon_0 }^{-1} \Upsilon^{2}\kappa^{-2}R/\of{k_B T},
\end{equation}
where $\epsilon$ is the permittivity of the solvent relative to vacuum,
$\epsilon_0$ is the permittivity of the vacuum, and $\kappa$ is 
the inverse Debye screening length. We have adopted 
the experimental values $\Upsilon=5.7\mu C/cm^2$ and $2R=555nm$
as given in Ref.~\onlinecite{Kurnaz}(b). The relative permittivity $\epsilon$ 
as function of the mass fraction \m[L] of lutidine
can be obtained by using the Clausius-Mossotti relation
\cite{Law:98}${}^{\text{b,}}$\cite{note_permittivity}{}. 
We have taken the values $\xi_0=0.25nm$ and $\tcb=310 K$ as typical
for water-lutidine mixtures (see, e.g., Ref.~\onlinecite{Gambassi-et:2009}).
In Ref.~\onlinecite{Kurnaz}(b) no precise value of the Debye screening length
$\kappa^{-1}$ is given; it is stated that 
$\kappa^{-1}\;\!\approx\of{7\ldots 10}nm$.
Therefore in \fref{fig:b2_comparison} we show
results for $\bsec$ calculated for several values of 
$\kappa^{-1}[nm]=5,7,10$.

Within their analysis, for temperatures far away from $\tcb$ 
(i.e., $\vert T-\tcb\vert \gtrsim3K$) the authors of Ref.~\onlinecite{Kurnaz}(b)
obtained values for $\bsec$ which would correspond to a system of hard 
spheres of a radius of $4200nm$. These extremely large, 
unexpected values (they actually used particles of diameter $555nm$) 
cannot be easily explained by a soft, repulsive interaction {\cite{Kurnaz}$^{\text{(b)}}$}. 
In order to circumvent this non-trivial effect, in 
\fref{fig:b2_comparison} we plot $\bsec$ normalized by the 
corresponding values $\of{\bsec}_{reg}$ far away from \tcb. For the 
normalization of the experimental data, we have used the
mean values corresponding to the three largest temperature
differences \abs{T-\tcb} which have been reported.
For $\of{\bsec}_{reg}$ in the theoretical curves we have used the
values as obtained by taking in \eref{eq:coll_potential} for $x>0$
only the soft, repulsive term into account
(i.e., taking $\zeta=\kappa\xi=0$), so that 
$\of{\bsec}_{reg}/v_{HS}= 5.28, 6.00 ,7.25$ for 
$\kappa^{-1}[nm]=5,7,10 $, respectively, where 
$v_{HS}=\of{4\pi/3}R^{3}$ is the volume of 
the hard spheres. 

Although we have adopted a rather simple form for the background
repulsive potential, for $\kappa^{-1}=5nm$
we obtain a fair agreement between the normalized experimental and
theoretical values. Yet due to large error bars
of the experimental data and large differences between not normalized
experimental and theoretical values of \bsec{} (the origin of which is
unclear) we do not claim that this value of $\kappa^{-1}$ is
the actual experimental one. Rather we consider
$\kappa^{-1}$ in that case as an effective fitting parameter which
accounts for differences between the experimentally
realized and the model background potential in order to match the model
with the experimental data. Furthermore, we have not checked the sensitivity 
of our results to uncertainties in the other parameters. 
Both from experiment and theory it is evident that the attraction close
to the critical temperature increases significantly,
resulting in a steep decrease of \bsec{} towards large negative
values. 

\begin{figure}
	\includegraphics{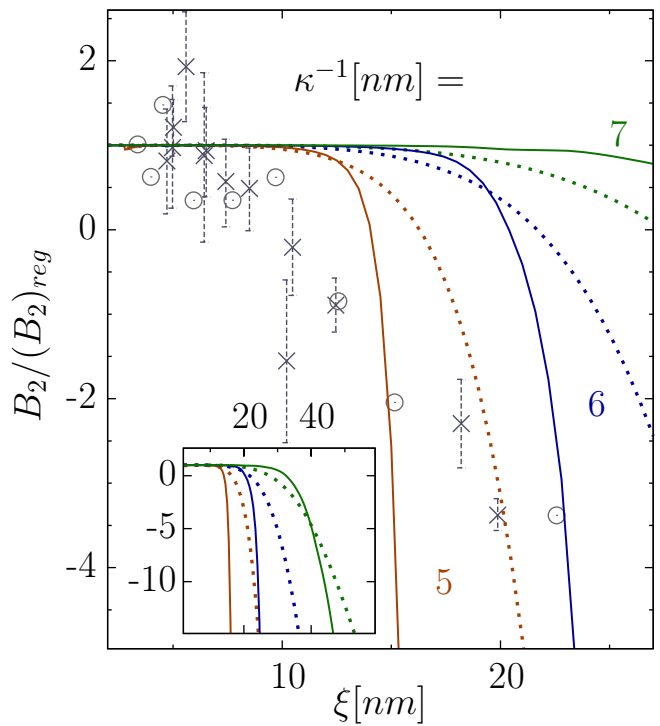}
	\caption{
	The variation of the second virial coefficient $\bsec$ 
	(\eref{eq:secvirial}) upon approaching the critical temperature 
	\tcb{} for an off-critical composition of the binary solvent. The
	same data in the inset, but as function of the correlation
	length $\xi=\xi_0\abs{1-T/\tcb}^{-\nu}$ of the solvent.
	The experimental values ($\times$) are taken from Fig.~2
	of Ref.~\onlinecite{Kurnaz}(b) and the three theoretical curves are obtained 
	by using the effective potential in \eref{eq:coll_potential}.
	The values $\kappa^{-1}[nm]=7,\,6,\text{ and } 5$ correspond
	to $m_0=-10.6,\,-13.5,\text{ and } -18.0$, respectively
	(\eref{eq:op_amplitude_modified}).
	For the lutidine mass fraction $\m[L]=0.35\gtrsim \m[L,c]$ the
	solvent is poor in the component preferred by the colloids. 
	Therefore the critical Casimir force is stronger than for 
	the critical composition. Thus $\bsec$ can be expected to 
	decreases more rapidly than for $\m[L]=\m[L,c]$.
	This is indeed the case for the theoretical curves. 
	For the experimental data of Ref.~\onlinecite{Kurnaz}(b) no dependence on \m[L] is seen;
	the dotted curves and the experimental data $\of{\circ}$ correspond to the
	critical composition as shown in \fref{fig:b2_comparison}. 
	$\bsec$ is normalized in the same way as in \fref{fig:b2_comparison}.
	The order parameter amplitude used here is
	$\mathcal{B}_{\m}=0.765$ (see \eref{eq:op_amplitude_modified}, 
	Ref.~[73] in Ref.~\onlinecite{MMD_part1}, and the text below Eq.~(27) 
	in Ref.~\onlinecite{MMD_part1}). By making use of the Clausius-Mossotti 
	relation \cite{Law:98}${}^{\text{b,}}$\cite{note_permittivity}{} we obtain $\epsilon=19.3$ (see \eref{eq:A_electric}).
	The values of the other parameters are the 
	same as in \fref{fig:b2_comparison}.}
	\label{fig:b2_comparison_off}
\end{figure}

\begin{figure}
	\includegraphics{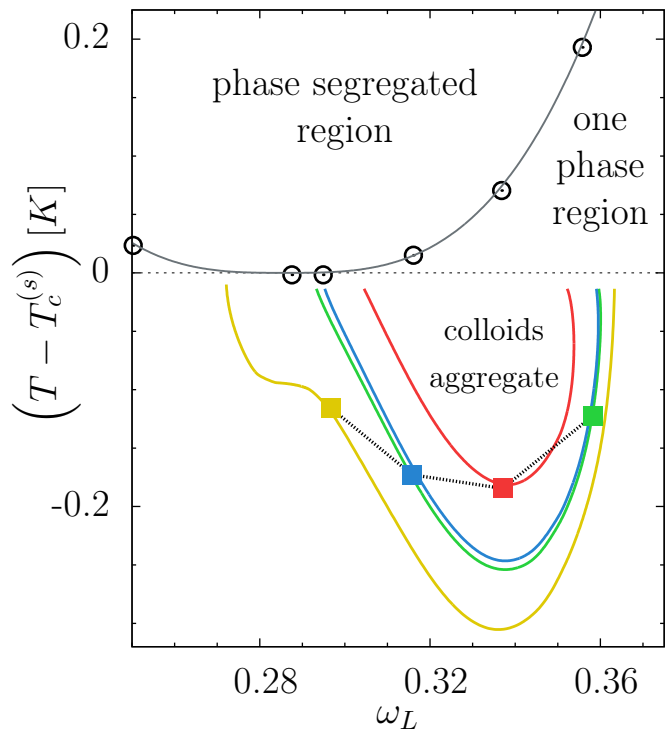}
	 \caption{ 
	 Comparison of the values of $\bsec$ (\eref{eq:secvirial})
	 at the thermodynamic state points of the solvent
	 for which immersed colloids aggregate.
	 The four experimentally obtained state points (squares;
	 the straight black dotted lines in between are a guide
	 to the eye) are taken from Fig.~1 (middle) of Ref.~\onlinecite{gallagher:92}(b)).
	 The isolines of constant $\bsec$ (full colored lines),
	 each belonging to one of the state points (squares), are
	 calculated within our model (\eref{eq:coll_potential}).
	 Each $\bsec$-isoline can capture some qualitative trends of
	 the possible shape of the line of the onset of aggregation,
	 but they cannot serve as quantitative indicators for the onset
	 of aggregation. For further details see the main text.
	 In order to obtain the theoretical values of
	 $\bsec$ we have taken $\xi_0=0.25nm$, $\kappa^{-1}=6nm$,
	 $\tcb=307K$, and $\m[L,c]=0.286$ and we have used the 
	 experimentally given values $2R=378nm$ and $\Upsilon=3.85 \mu C/cm^{2}$.
	 The repulsion strength $A$ is obtained from \eref{eq:A_electric}
	 with $\epsilon$ depending, according to the Clausius-Mossotti 
	 relation \cite{Law:98}${}^{\text{b,}}$\cite{note_permittivity}{}, 
         on the mass fraction \m[L] of lutidine.
	 The experimentally determined coexistence points (circles)
	 on the binodal of phase segregation with a lower critical
	 point agree well with the relation
	 $\abs{\m[L]-\m[L,c]}=\mathcal{B}_{\m}\abs{t}^{\beta}$, where
	 $\beta=0.3256$ and $\mathcal{B}_{\m}=0.765$ (gray line); 
	 in accordance with a lower critical point
	 the reduced temperature is $t=\of{\tcb-T}/\tcb$.
	 }
	\label{fig:aggrpoints_comparison}
\end{figure}

In Ref.~\onlinecite{Kurnaz}(b) the variation of $\bsec$ as function of $\abs{T-\tcb}$
is reported also for a mass fraction $\m[L]$ of lutidine larger than 
its critical value $\m[L,c]\simeq 0.287$ (\fref{fig:b2_comparison_off}). (This value of
$\m[L,c] = 0.287$ is quoted in Ref.~\onlinecite{Kurnaz}(b); it differs slightly from 
the value $\m[L,c]= 0.286$ quoted in Ref.~\onlinecite{Beysens-et:1985} and 
as obtained from fitting the two-phase coexistence curve determined experimentally
in Refs.~\onlinecite{Beysens-et:1985,gallagher:92} to the functional
form $\abs{\m[L]-\m[L,c]}=\mathcal{B}_{\m}\abs{t}^{\beta}$
with $\beta=0.3256$; see \fref{fig:aggrpoints_comparison}.
In other papers values are quoted within the range $\m[L,c]=0.28 \ldots 0.29$; see, e.g.,
Refs.~\onlinecite{Beysens-et:1985,Gambassi-et:2009, Jayalakshmi-et:1994,Gurfein-et:1989}.)
In general, for off-critical compositions the strength and the range of
the critical Casimir force results from the interplay of the two bulk
correlation lengths $\xi\sim \abs{t}^{-\nu}$ and 
$\xi^{\of{h}}\sim \abs{\hb}^{-\nu/\of{\beta\delta}}$ (see 
\eref{eq:coll_potential0} and the discussion thereafter). 
For $\m[L]=0.35$ the binary solvent is poor in the component preferred 
by the kind of colloids used (i.e., water). In such a case the interplay 
of $\xi$ and $\xi^{(h)}$ is rather complex due to the occurrence of 
capillary condensation which leads to bridging between two colloids. In particular,
at the same temperature the CCF can become stronger upon moving away 
from the critical composition. But at off-critical compositions it is in general shorter
ranged than at the critical composition.
Moreover, at off-critical compositions the CCF emerges as function 
of temperature only closer to the critical temperature as compared with the 
corresponding thermodynamic path at the critical composition. 
In \fref{fig:b2_comparison_off} we compare the reduced experimental data for
\bsec{} corresponding to the mass fraction $\m[L]=0.35>\m[L,c]=0.287$ 
with the theoretical curves obtained within our effective model; 
for $\m[L]=0.35$ one has $\of{\bsec}_{reg}/v_{HS}=5.30,5.65,6.03$ 
for $\kappa^{-1}[nm]=5,6,7$, respectively.
As can be inferred from \fref{fig:b2_comparison_off} for $\kappa^{-1}=5nm$ 
the total attraction (within our effective model), in terms of
\bsec{}, is stronger for $\m[L]=0.35$ than for the critical composition
\m[L,c], i.e., for the same value of $\xi$, \bsec{} is more negative. For
the theoretical curve corresponding to $\kappa^{-1}=7nm$ the aforementioned 
interplay of $\xi$ and $\xi^{\of{h}}$ manifests itself more
clearly. For the off-critical composition the attraction due to the 
CCFs sets in only for larger values of $\xi$ but becomes more rapidly
stronger with increasing $\xi$ so that, for large values of
$\xi$, \bsec{} is more negative for $\m[L]=0.35$ than for $\m[L,c]$.
At the critical composition, \bsec{} starts to decrease (due to the
attractive CCFs) already for smaller values of $\xi$ than for $\m[L]=0.35$. 

For the experimental data (within the large error bars and the
reported temperature range) no dependence of the value of \bsec{}
as function of $\abs{T-\tcb}$ on the composition is observed.
(Reference~\onlinecite{Kurnaz}(b) provides the values of \bsec{} as function of
$\abs{T-T_{cx}^{(s)}}$ but lacks an absolute value
for the coexistence temperature $T_{cx}^{(s)}$; $T_{cx}^{(s)}$ is
the temperature at which the solvent undergoes a first-order phase 
segregation for $\m[L]=0.35\neq \m[L,c]$; see Fig.~1(a) in
Ref.~\onlinecite{MMD_part1}. We take
$T_{cx}^{(s)}-\tcb=0.145 K$ as obtained \cite{MMD_part1} from
$\abs{\m[L]-\m[L,c]}=\mathcal{B}_{\m}\abs{t}^{\beta}$ with
$\mathcal{B}_{\m}=0.765$.)
This finding is in contrast to the behavior of the measured effective particle-wall
interaction potentials of Ref.~\onlinecite{Nellen:2011}, for which a dependence 
on the composition of the water-lutidine mixture has been observed.
At present it remains unclear whether the dependence of $\bsec$
on $\m[L]$ could not be resolved experimentally or whether the 
data from these experiments happen to lie in that crossover regime
in which the curves $\bsec\of{\tcb-T}$ happen to be close to each
other for the critical and the off-critical 
mixture (see the inset in \fref{fig:b2_comparison_off}).

For the same system the authors of Ref.~\onlinecite{Kurnaz}(b) used for the
measurement of $\bsec$, in Ref.~\onlinecite{gallagher:92}(b) the aggregation phenomena
of immersed colloids have been studied. We have used the 
corresponding given experimental values of the parameters, and especially
\eref{eq:A_electric} for the repulsion strength parameter $A$, in order to
investigate these state points within our model.

First, it is worthwhile to mention that for the parameters
corresponding to these experiments, the repulsive barrier for small 
values of $x=\kappa D$ is of the order of tens of $k_B T$. 
Therefore, it seems that the observed aggregation is due to the very 
deep secondary minimum emerging at a certain distance $D_{min}$.
This is a different mechanism than the one discussed
in Sect.~\ref{ssc:res_stability}, where
the aggregation due to the disappearance
of the repulsive barrier at short distances $D$ has been considered.
While in the latter case the colloids would
stick together with their surfaces at contact, 
in the former case they are close but not at contact. 
With $\kappa D_{min}\approx 1$ and $s=\kappa R \approx 10\ldots100$, 
from the effective potential given in \eref{eq:coll_potential}
it follows that the colloids are a surface-to-surface distance
$D/R\approx 10^{-1}\ldots 10^{-2}$ apart. In Ref.~\onlinecite{gallagher:92}(b)
the mean surface-to-surface distance has not been estimated, but in
experiments performed for a similar system (see Ref.~\onlinecite{Gurfein-et:1989})
it has been found that in the flocculated and sedimented
aggregates the average interparticle distance is about three times
the mean radius of the polydisperse colloids. This is in the stark contrast with
our theoretical estimates. Other
experimental reports do not specify the average interparticle
distance occurring in the aggregates. In Ref.~\onlinecite{grull:97} it is 
stated, that the sediment is less dense than expected for a tight 
packing of spheres and in Ref.~\onlinecite{Beysens-et:1999} the aggregates 
are characterized as being compact, not fractal. The authors of
Ref.~\onlinecite{Koehler-et:1997} argue that for their system
the aggregates are indeed crystalline. 
The possibility, that the flocculated colloidal particles  form a bona fide,
thermodynamically stable, liquid-like or crystalline phase cannot be
addressed within the effective, one-component approach used here, because 
within this approach the reliable prediction of the occurrence of a phase
transition can be made only if the colloidal number density is sufficiently 
large (see Ref.~\onlinecite{MMD_part1}, especially, Subsect.~III~B~3), whereas in
the experiments referred to above the packing fraction is 
as small as $\eta\; \approx \;10^{-6}\ldots10^{-3}$.

Next, we address the issue whether a relation can
be established which connects the onset of the reversible
aggregation to a threshold of a specific quantity
which is both theoretically and experimentally accessible.
To do so we follow Ref.~\onlinecite{Kurnaz} and analyze
whether the values of $\bsec$ corresponding
to thermodynamic state points of the solvent, at which the aggregation
of colloids sets in, are similar.
For such state points, as determined experimentally
and reported in Fig.~1 (middle) of Ref.~\onlinecite{gallagher:92}(b), we have calculated
the corresponding values of $\bsec$ within our model
for four values of $\omega_L$ yielding \cite{note_b2} 
$\bsec/\of{\frac{4\pi}{3}R^3}=$ $-7400$ (red square), $-560$ (blue), $-200$ (green),
and $-76.0$ (yellow) (Fig.~\ref{fig:aggrpoints_comparison}).
In Fig.~\ref{fig:aggrpoints_comparison} the isolines of constant
$\bsec$ for these values of $\bsec$ are shown 
(using the same color code). Clearly the four curves do not 
coincide to form a single curve with the four squares on it which 
would support the hypothesis that aggregation sets in at a threshold 
of value of $\bsec$. However, one has to keep in mind that the CCFs
strongly depend on the temperature. This dependence is amplified 
within the resulting value of $\bsec$ due to its exponential
dependence on the potential. Therefore the value of $\bsec$ may vary
significantly even within a narrow temperature interval (compare
Figs.~\ref{fig:b2_comparison} and \ref{fig:b2_comparison_off}).
Moreover, small uncertainties in the measured temperature, at which
the aggregation sets in, result in a large variation of $\bsec$.
Each $\bsec$-isoline reproduces some of the qualitative trends of the
experimental data points for aggregation onset, but there is no
satisfactory quantitative agreement. Either the isoline is too narrow 
(see the red curve which cannot capture any of the yellow, blue, or
green squares) or the isoline extends to too large values of 
$\abs{T-\tcb}$ (see, e.g., the yellow curve which misses the blue, red,
and green squares). Besides the possibility that 
$\bsec$ is indeed not the same at the various state points at which
coagulation sets in, there are further reasons
which could explain this disagreement. One reason could be the
simplified choice for the background potential which
we have used. Furthermore, there are several experimental 
parameters for which we could use only estimates. 
Within their analysis, the authors of Ref.~\onlinecite{gallagher:92}(a) 
conclude that for calculating the electrostatic potentials one should
not use the bare surface charge density as given 
by the manufacturer. Rather, the dissociation of the colloidal 
surface in the solvent has to be taken into account.
In our analysis we took the nominal value for the surface charge
due to the lack of more detailed data. The
authors of Ref.~\onlinecite{narayanan:93} point this out, too,
and reveal the role of impurities as another important factor. They
determined the aggregation line in ``fresh'' mixtures and in the same 
mixture but one or several days later. While the aggregation
phenomenon as such was reproducible, the values of the thermodynamic 
variables of the solvent at which the aggregation sets in varied strongly
within one and the same sample.

In the same reference, a dependence of the aggregation line on 
the volume fraction $\eta$ of colloids is observed, whereas in 
Ref.~\onlinecite{Koehler-et:1997} it is stated that the shape of the 
region in thermodynamic space where aggregates are forming does 
neither depend on the colloidal concentration nor on the colloidal 
radius. In Ref.~\onlinecite{gallagher:92}(b) a dependence of the time scale of the
aggregation process on $\eta$ is reported. Our analysis does not
capture such a dependence.

In conclusion, our results suggest that the critical Casimir force
plays an important role in order to explain the aggregation phenomenon
in this region of the phase diagram of the solvent. 
Yet, in order to be able to determine all relevant
mechanisms and forces and in order to obtain 
quantitative agreement between the experiments and
the theoretical results, for both further efforts
are needed to provide more precise values of the 
relevant quantities.

Finally, in Fig.~\ref{fig:aggrpoints_comparison} experimental
data for the binodal of the water-lutidine 
mixture are shown, which nicely 
agree with the functional form
$\m[L]-\m[L,c]=\pm\mathcal{B}_{\m}\abs{t}^{\beta}$
(gray line), where $\beta=0.3256$ and $\mathcal{B}_{\m}=0.765$. 
This shows that our description, which is symmetric w.r.t.
$\m[L]=\m[L,c]$, is sufficient for the phase space region of
interest here.

\section{Summary \label{sec:summary}}

We have studied aggregation phenomena of monodisperse colloidal
suspensions with near-critical solvents. In such solvents, 
driven by the diverging bulk correlation length
$\xi\left(t=\of{\tcb-T}/\tcb\to0^{\pm}\right)=\xi_0^{\pm}\abs{t}^{-\nu}$,
so-called critical Casimir forces act on the colloidal 
particles, in addition to regular background forces. 
In order to obtain our results, we have adopted a description
in terms of an effective one-component system.
Within this approach the presence of the solvent is taken into account 
via contributions to the effective pair potential of the colloids.
We have considered an effective pair potential 
for the colloids (with radius $R$) which consists of the critical 
Casimir contribution plus a regular background potential. For the latter 
we have employed a generic, softly repulsive contribution with strength 
$A$ and acting on a length scale $\kappa^{-1}$.
We have focused on dilute suspensions with low colloidal number densities 
$\rho$, for which the approximation of pairwise additive critical
Casimir forces (CCFs) is valid. In addition, we have applied the Derjaguin
approximation (\eref{eq:sfpot_derjaguin}), within which the pair 
potential between two spheres can be expressed in terms of
the scaling function of the CCF in a slab.
In order to obtain access to the dependence of the universal 
scaling function for the CCF on the 
scaling variable $\Sigma$ associated with the bulk ordering field 
$\hb$ in spatial dimension $d=3$ we have employed an approximation 
based on a combination of both Monte Carlo simulation data 
and of mean-field theory results (\eref{eq:sfp_h_approx}).
The resulting effective pair potential 
is given in \eref{eq:coll_potential} 
and various examples are plotted in \fref{fig:coll_potential}. 
On this basis we have obtained the following main results:
\begin{enumerate}
\item
	We have calculated the stability ratio
	$\w\sim t_{pair}/t_{diff}$ (Eqs.~\eqref{eq:stabilityratio} and
	\eqref{eq:stability_timescales}), which is a measure of the
	stability of a colloidal suspension. The time scales $t_{pair}$ 
	and $t_{diff}$ are related to the pairing of two colloids
	and to the diffusion of a single particle, respectively.
	Rapid coagulation is expected to set in for $\w<1$.
	In \fref{fig:stability_diagram} the contour
	lines $\w=1$ are shown for various values of
	$\zeta=\kappa\xi\sim\abs{t}^{-\nu}$
	and $s=\kappa R$ in the space spanned by the
	parameter $a=A/s$ and by the scaling variable $\Sigma$. 
	For the case of a binary liquid mixture $\Sigma$
	is related to the difference of the chemical potentials of
	the two species. This shows that the CCF
	can lead to rapid coagulation. The coagulation region extends
	to larger values of $a$ for $\Sigma\lesssim0$ because the CCFs
	are stronger for compositions of the solvent slightly poor
	in the component preferred by the colloids.
\item
	The variation of $\w$ with temperature at a fixed solvent
	composition, as it is typically realized in experiments, 
	is shown in \fref{fig:stab_temperature}. We find that \w{}
	rapidly decreases to $\w<1$ upon approaching the critical
	temperature via the disordered phase.
	This implies, that the aggregation should set in within a
	narrow temperature interval.
\item
	Concerning the structure of the colloidal suspension,
	we have solved numerically the Ornstein-Zernicke
	equation (\eref{eq:oz}) using the Percus-Yevick and the
	hypernetted-chain closure (Eqs.~\eqref{eq:py_k} and
	\eqref{eq:hnc_k}, respectively).
	In Figs.~\ref{fig:structure_simpleMin} and
	\ref{fig:structure_sevExtrema} we show the
	radial distribution function $g\of{r}$ for two
	different parameter sets \of{a,s,m_0} and various
	values of $\zeta$, where $m_0^{-1}\sim c_a-c^{(s)}_{a,c}$ is
	related to the composition of the solvent
	(\eref{eq:op_amplitude_modified}). 
	Close to the surface of a colloidal
	particle ($r\gtrsim2R$) a peak develops for
	increasing values of $\zeta$, reflecting the increasing attraction
	due to the CCF. This indicates the presence of a
	pronounced short-ranged order and that the formation of
	colloidal dimers is favored.
\item
	In Figs.~\ref{fig:b2_comparison} and
	\ref{fig:b2_comparison_off}, for the critical and for an
	off-critical composition, respectively, we compare the
	variation of the second virial coefficient $\bsec$
	(\eref{eq:secvirial}) upon increasing
	$\xi=\xi_0\abs{t}^{-\nu}$, as obtained 
	within our model, with experimental
	data {\cite{Kurnaz}$^{\text{(b)}}$}. For both, the theoretical results and the
	experimental data, 
	for increasing $\xi$ the virial coefficient \bsec{} decreases  
	towards large negative values. Due to extremely large
	experimental values of \abs{\bsec{}} also far away from \tcb{},
	the origin of which is not clear, the quality of the
	quantitative comparison is limited. 
	For the off-critical composition, the CCF can become stronger  
	than at the critical composition but it has in general a shorter range
	and occurs closer to the critical temperature of the solvent. 
	Therefore, for the same value of $\xi$, the total attraction 
	due to the CCFs can either be smaller or larger for an off-critical 
	composition compared with the critical one. Accordingly, as can 
	be seen in \fref{fig:b2_comparison_off}, \bsec{} can be less or more
	negative, respectively. For the experimental data of 
	Ref.~\onlinecite{Kurnaz}(b) there is no visible dependence 
	on the composition of the solvent, 
	which could be due to this crossover behavior.
  \item
	We have addressed the issue whether a relationship
	can be established between the onset of aggregation
	and a specific quantity which is accessible both
	theoretically and experimentally. To this end, within 
	our model we have calculated $\bsec$ at state points
	for which in experiments aggregation sets in. The
	corresponding $\bsec$-isolines
	capture qualitatively the main trends of the aggregation
	line, but no satisfactory quantitative agreement has been
	achieved (see Fig.~\ref{fig:aggrpoints_comparison}). We
	have discussed various possible explanations
	for this discrepancy.
\end{enumerate}

To conclude, our results show that CCFs, which can be easily 
controlled by temperature $T$ and the strength and range of which 
can be varied by changing the bulk ordering field \hb{}, can lead to 
aggregation in colloidal suspensions with near-critical solvents.
Within the approach of using an effective potential,
we have identified the ranges of values for the background repulsive
potential and the values of the scaling variables associated with the 
critical solvent, for which the colloids aggregate.

\begin{acknowledgments} 
We acknowledge S. Buzzaccaro and R. Piazza for informing us about their 
experimental results in Ref.~\onlinecite{Buzzaccaro-et:2010} before publication. 
We thank R. Evans for his generous interest in our work and 
the stimulating and enlightening discussions we enjoyed to 
have with him.
\end{acknowledgments}


\end{document}